\documentclass[aps,prl]{revtex4}

\usepackage{graphicx}

\begin{document}


\title{Predictive dynamical and stochastic systems}


\author{Toru Ohira} 
\email[]{ohira@csl.sony.co.jp}
\homepage[]{www.csl.sony.co.jp/person/ohira}

\affiliation{Sony Computer Science Laboratories, Inc., Tokyo, Japan 
141-0022}

\begin{abstract}
We study a system whose dynamics are governed by predictions of its future states. 
A general formalism and concrete examples are presented. We find that the dynamical characteristics depend on
how to shape the predictions as well as on how far ahead in time to make them. We also report that noise can induce
oscillatory behavior, which we call ``predictive stochastic resonance".
\end{abstract}

\maketitle


\section{Introduction}
Predictive behaviors are common in our everyday activities. A few examples include the timing of braking when driving a car, 
catching a ball, trading stock, and shaping population control policies.                                                                                                                                                                                                                                                                                                             
One of the main principles of normal dynamics is that the past and present decide the future. Most physical theories have
been founded on this principle. It should be noted, however, that the idea of identifying a positron going forward in time
with an electron ``coming back from the future" has brought new insight to elementary particle physics \cite{feynman1961}. 
Given the common occurrence of making predictions, a consideration of the
theoretical pictures of systems whose dynamics are explicitly
governed by predictions or estimations of future states may be constructive.
The main theme of this letter is to propose predictive dynamical systems
by presenting concrete examples. The behavior of such dynamical systems depends on how the predictions are made and 
on how far in advance they are made. We also report on that noise can induce
oscillatory behavior, which we call ``predictive stochastic resonance.''

\section{Model and Analysis}

We start with the general differential equation of predictive dynamical systems, given by
\begin{equation}
{dx(t) \over dt} = F(x(t),\bar{x}(\bar{t})).
\end{equation}
Here, $x(t)$ is the dynamical variable, and $F(x)$ is the dynamical function. $\bar{x}(\bar{t})$ is a prediction of $x$ at a future time $t<\bar{t}$.
This dynamical equation implies that the rate of change of $x(t)$ depends not only on its current state, but also on the predicted future state $\bar{x}(\bar{t})$ 
through the dynamical function $F$.
Naturally, when $\bar{t}=t$, it reduces to the normal dynamical equation. In this letter, we discuss the class of equations of the form
\begin{equation}
{dx(t) \over dt} = -\alpha x(t) + f(\bar{x}(\bar{t}))
\label{pde}
\end{equation}
with a constant $\alpha > 0$ for comparison with the corresponding delayed dynamical equations
\cite{mackeyglass1977,cookegrossman1982,glassmackey1988,milton1996}.

There is a variety of ways we can choose $\bar{t}$ and the prediction $\bar{x}(\bar{t})$. In this letter, we set $\bar{t} = t + \eta$ with a 
parameter $\eta$, which we call the ``advance." In other words, the dynamics are governed by the predicted state of the dynamical variable
$x$ at a fixed interval $\eta$ in the future. To predict $x$, we consider two cases. The first case is to extrapolate the dynamics for
the duration of the advance $\eta$ with normal ($\eta=0$) dynamics, so that
\begin{equation}
\bar{x}(\bar{t}=t+\eta) = \int_{t}^{t+\eta}{F(x(s))}ds + x(t).
\end{equation}
We call this the ``extrapolate prediction." The second case is to assume that the current rate of change of $x$ continues for the duration of the advance.
This case is termed the ``fixed rate prediction" and is given by 
\begin{equation}
\bar{x}(\bar{t}=t+\eta) =\eta {dx(t) \over dt}+ x(t).
\end{equation}

We examine how these different predictions, together with the value of $\eta$ affect the dynamics.  We conducted our investigation by computer simulation. 
To avoid ambiguity and for simplicity, we consider time-discretized map 
dynamical models, which incorporate the above--mentioned general properties of the predictive dynamical equations.
The general form of a predictive dynamical map is given as follows.
\begin{equation}
x(t+1) = (1-\mu)x(t) + f(\bar{x}(t+\eta))
\end{equation}
where $\mu$ is a rate constant. The extrapolate prediction can be obtained by iterating the corresponding normal map ($\eta=0$) for the duration of $\eta$.
The fixed rate prediction is obtained by setting $\bar{x}(\bar{t}=t+\eta) = \eta (x(t)-x(t-1)) + x(t)$.

The first model we consider is a ``sigmoid" map (fig. 1(a)) with

\begin{equation}
f(x) = {2 \over {1 + e^{-\beta x}}} - 1,
\end{equation}

\begin{figure}
\includegraphics[width=8cm]{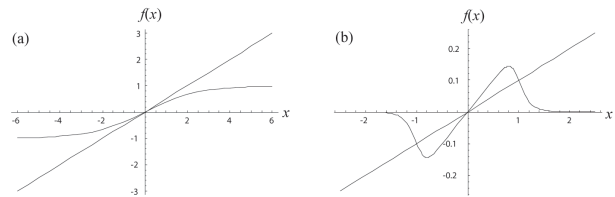}
\caption{
Dynamical functions $f(x)$ with parameters as examples of the simulations discussed in this letter. (a) Sigmoid function with $\beta = 0.8$. 
The straight line has a slope of $\alpha = 0.5$. (b) Mackey-Glass function with $\beta = 0.8$ and $n = 10$. The straight line has 
slope of $\alpha = 0.1$.
}
\end{figure}
\begin{figure}
\includegraphics[width=9cm]{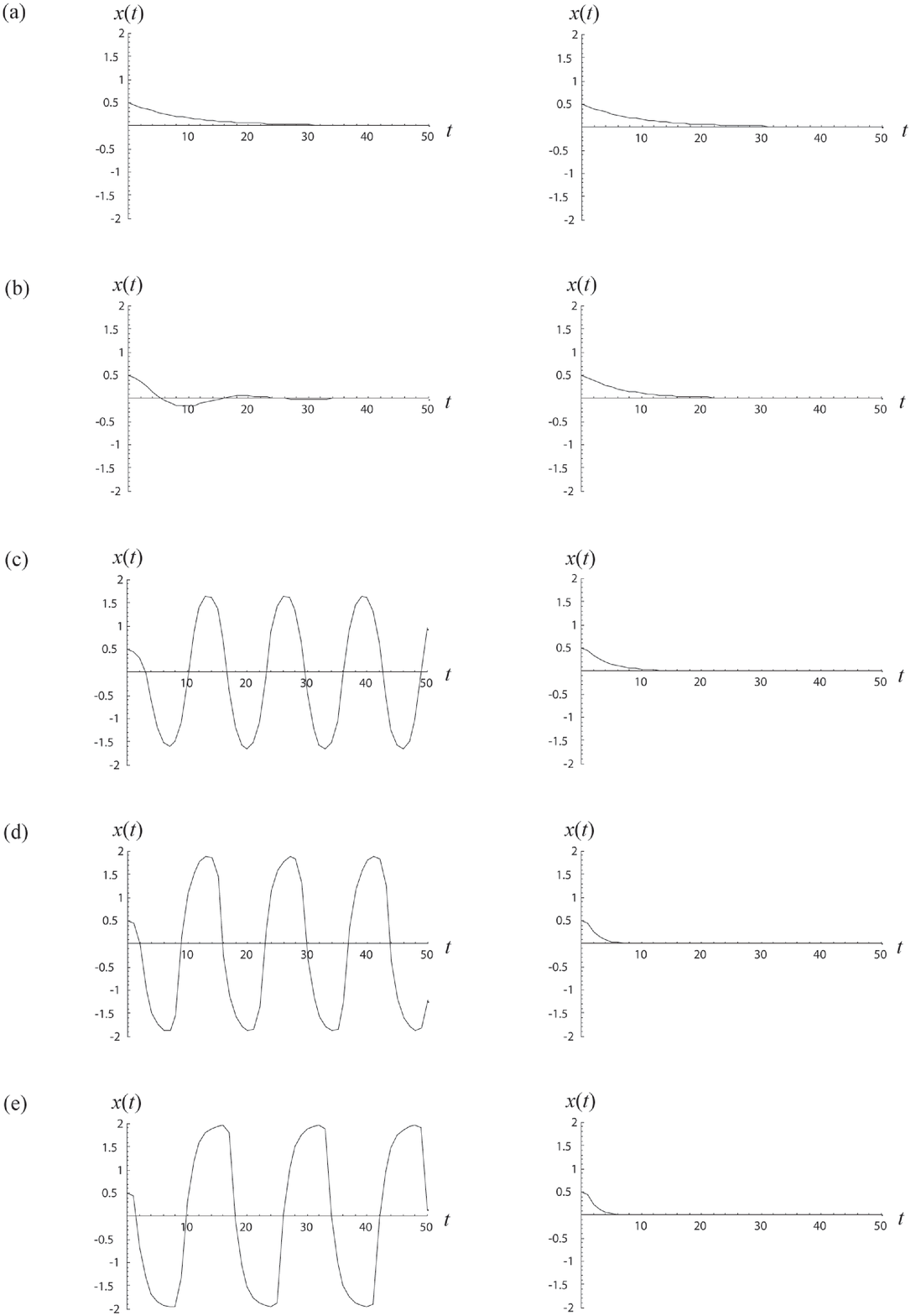}
\caption{
Examples of fixed rate (left column) and extrapolate (right column) predictions for 
a sigmoid map with $\mu=0.5$ and $\beta = 0.8$. The initial condition is $x(0) = 0.5$, and for predictive 
dynamics, $x(1) = (1-\mu) x(0) + f(x(0))$.
The values of the advance $\eta$ are (a) 0, (b) 2, (c) 5, (d) 20, and (e) 80.
}
\end{figure}

This function is often used in the context of neural network modeling \cite{cowansharp1989,hertz1991}. We simulated this model with the extrapolate and with the fixed rate predictions. Some examples are shown in Fig. 2. In these examples, we have set the 
parameters $\mu$ and $\beta$ to the same values in both prediction schemes. Given this parameter set, the origin $x=0$ is a stable fixed point when there is no advance, i.e., $\eta=0$.
In the case of the extrapolate prediction, this property is kept even when $\eta$ is increased. 
The situation is quite different for the
case of fixed rate predictions. Here, an increasing $\eta$ breaks the stability of the origin, and periodic behavior arises.

A similar comparison can be made by setting the dynamical function to the ``Mackey-Glass" map (Fig. 1(b)):
\begin{equation}
f(x) ={ {\beta x} \over {1 + x^{n}}}.
\end{equation}
This function was first proposed for modeling the cell reproduction process, and it is known to induce chaotic behavior with a large delay \cite{mackeyglass1977}.
Figure 3 shows the examples of the computer simulations we conducted. We can see that even though the extrapolate prediction 
does not change the stability of the fixed point with an increasing advance, the fixed rate prediction case gives rise to complex dynamical behavior.

\begin{figure}
\includegraphics[width=9cm]{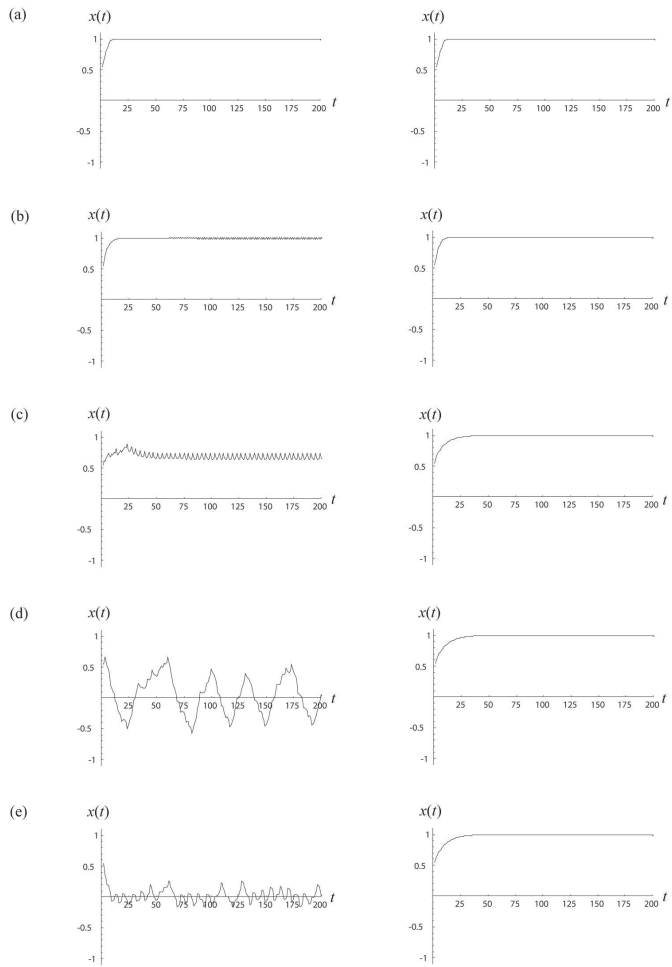}
\caption{
Examples of fixed rate (left column) and  extrapolate (right column) predictions for 
a Mackey-Glass map with $\mu = 0.5$, $\beta = 0.8$, and $n = 10$.
The initial condition is $x(0) = 0.5$, and for predictive 
dynamics, $x(1) = (1-\mu) x(0) + f(x(0))$.
The values of the advance $\eta$ are (a) 0, (b) 2, (c) 8, (d) 10, and (e) 20.
}
\end{figure}

Now, we turn our attention to the effect of noise. The main motivation here stems from the fact that a combination of prediction dynamics
and noise could lead to behaviors similar to ``stochastic resonance" \cite{wiesenfeld-moss95,bulsara96,gammaitoni98}. Stochastic resonance
has been studied in variety of fields \cite{mcnamara88,longtin-moss91,collins1995,chapeau2003,lee2003}. In particular, we have proposed  a variation
of stochastic resonance, called delayed stochastic resonance, which arises from a combination
of noise with delayed dynamics \cite{ohira-sato,tsimring01,masoller}. 
We can infer from such an effect that a similar resonance
behavior could be found in fluctuations in predictive dynamics. We shall see that
this is indeed the case.

The model we discuss is an extension of the sigmoid map model with fixed rate prediction.
\begin{eqnarray}
x(t+1) & = & (1-\mu)x(t) + f(\bar{x}(t+\eta)) + \sigma \xi(t), \\
\bar{x}(\bar{t}=t+\eta) & = & \eta (x(t)-x(t-1)) + x(t),\\
f(x) & = & {2 \over {1 + e^{-\beta x}}} - 1.
\end{eqnarray}
Here, we have added a time--uncorrelated noise term $\xi$. $\xi$ takes a value between 
$(-1, 1)$ with a uniform probability, and the parameter $\sigma$ controls its 
``width" or ``strength". We have studied the behavior of this model with various
parameter sets. A resonance type of behavior occurs for a
parameter set whereby the dynamics is a simple monotone approach to a fixed point
without noise (Figure 4). Without noise or with small
noise, the behavior is a monotone approach to one of the fixed points or fluctuation
around a point. As the noise width increases, we begin to see oscillatory 
behavior. They are seen both in the time series of $x$ and in the associated power
spectrum. With too much strength in noise, however, the oscillation begins to loose
its regularity. The signal--to--noise ratio at the peak of the power spectrum is
used as an indication of this resonant behavior(Figure 4(f)).
The signal--to--noise ratio reaches a maximum with a ``tuned" noise strength.
We note that as in  delayed stochastic resonance, there is no external oscillatory
forces or signals present in the system. The oscillation is due to the combination
of the fixed rate prediction with appropriately chosen added noise. In this sense,
it is a new kind of stochastic resonance.

\begin{figure}
\includegraphics[width=8cm]{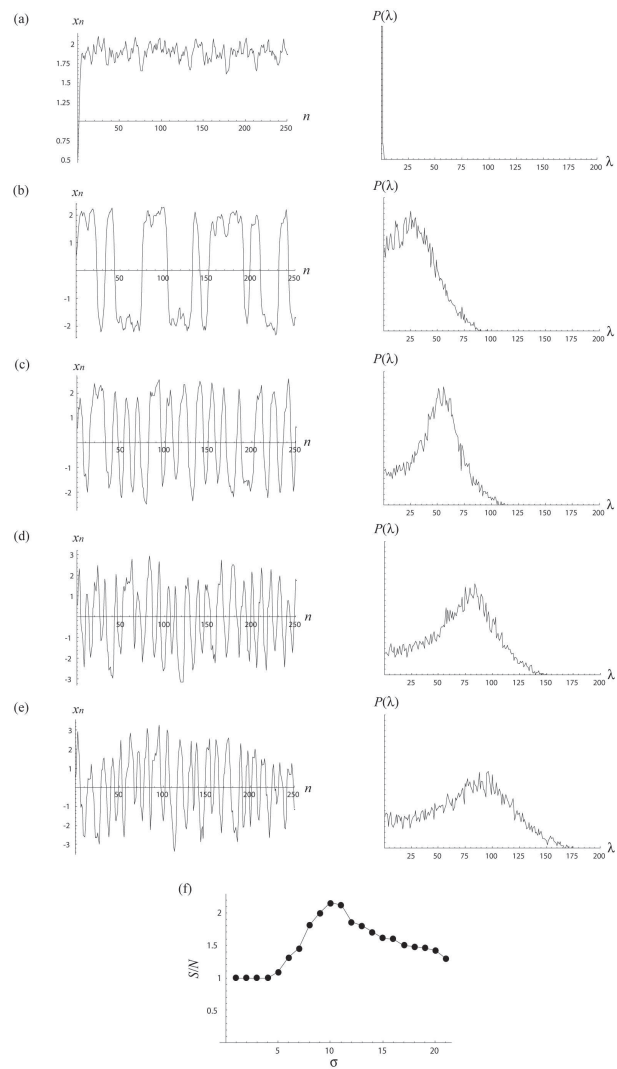}
\caption{Dynamics (left) and power spectrum (right) of predictive dynamical model with
sigmoid map. This is an example of dynamics and associated power 
spectrum simulatied using the model of Eq. (8--10) with a variable noise 
strength $\sigma$.  The 
parameters are $\mu = 0.5$, $\beta = 2$, $\eta=4$, and the initial condition is $x(0) = 0.5$, and $x(1) = (1-\mu) x(0) + f(x(0))$. 
The
noise strengths are (a) $\sigma=0.1$, (b) $\sigma=0.25$, (c) $\sigma=0.4$, (d) 
$\sigma=0.75$, and (e) $\sigma=1.0$. The simulation is performed for 
$L=1024$ steps and 50 averages are taken for the power spectrum. The unit of frequency $\lambda$ is ${1 \over L}$, 
and the power $P(\lambda)$ is in arbitrary units. (f) 
The signal--to--noise ratio ${S/N}$ at the peak height as a function of the noise strength $\sigma$. 
}
\end{figure}

\section{Discussion}

Now, we would like to discuss a couple of issues related to the results of the predictive dynamical models.
First, let us examine the difference in dynamical behavior between the extrapolate and the fixed rate predictions. 
Analytically, we can expand eq. (\ref{pde}) around the fixed point to examine its stability. 
In particular, we can obtain the following from linear stability analysis:
\begin{equation}
(1-\beta\eta){dz(t) \over dt} = (-\alpha+\beta)z(t), \quad z \equiv x - x^{*}, \quad \beta = {df \over dx}{|_{x=x^{*}}},
\end{equation}
where $x^{*}$ is the fixed point. For the case of the fixed rate 
prediction,  we can see that the advance, $\eta$, can switch the stability.  In the extrapolate prediction, on the other hand, 
the stability is not affected by the advance, provided that the corresponding normal dynamics monotonically approach
the stable fixed point. (The details of this stability analysis will be discussed elsewhere.)
Qualitatively, we can argue that the fixed rate predictions tend to ``overshoot" in
comparison with the extrapolation, leading to destabilization of the fixed point with a larger advance, $\eta$. 
Higher order analysis and other analytical tools need to be developed to understand these types of equations 
and capture their dynamical behaviors.

Second, we can compare our results with the case of delayed dynamics. In the case of delayed dynamics, we need to decide on the initial function and delay. Analogously, in predictive dynamics,
the prediction scheme and advance need to be specified. Common to the delayed and predictive dynamical systems, both factors affect the nature
of the dynamics. Also, as we have seen, we can have ``predictive stochastic resonance" in
a similar manner as delayed stochastic resonance.

Finally, in the same way that we have considered random walks with a delay (delayed random walks)\cite{ohiramilton1995,ohirayamane2000}, random walks with a prediction (predictive random walk) can also be considered.
Even with a small delay, delayed random walks give rather complex analytical expressions for statistical quantities, such as variance.  Analogously, the
analysis of predictive random walks is not straightforward. Mathematically, one may argue that these predictive dynamical and stochastic models can
be cast into the framework of normal non-linear dynamical systems, as, after all,  predictions are based on current and past states. Indeed, we can apply
linear stability analysis to gain a partial understanding. 
However, in theoretical modeling, particularly for such fields as physiological controls, economical or social behaviors, and  ecological studies, explicitly taking future predictions into account may be useful. 
For example, a recent study on humans performing stick balancing tasks on a fingertip has revealed that the 
corrective motion of the stick is frequently shorter than the neuro-physiological response time \cite{cabreramilton2002,cabreramilton2004a,cabreramilton2004b}. Also, there is an indication that external physical or 
intentional fluctuations can lead to better balancing\cite{ohirahosaka,hosakaohira,miltonohira}.   This is a task where both the feedback delay and predictions are intricately
mixed. Models based on delayed dynamics have been proposed, but such models may be further developed by using 
 predictive factors to investigate the experimental results. Models for this and other concrete applications have yet to be constructed 
and further analysis of predictive dynamical systems seems to be warranted.


\end{document}